\documentclass[twocolumn]{aastex631}

\usepackage{graphicx}	
\usepackage{amsmath}	
\usepackage{amssymb}	
\usepackage{bm}		

\usepackage[T1]{fontenc}
\usepackage{ae,aecompl}

\usepackage{newtxtext,newtxmath}

\begin{document}

\title[Dust Morphology Methodology]{The Morphology of Asteroidal Dust Around White Dwarf Stars: Optical and Near-infrared Pulsations in G29-38}

\author[0000-0002-5775-2866]{Ted von Hippel}
\altaffiliation{Visiting Astronomer at the Infrared Telescope Facility, which is operated by the University of Hawaii under contract 80HQTR19D0030 with the National Aeronautics and Space Administration.}
\affiliation{Department of Physical Sciences and SARA, Embry-Riddle Aeronautical University, Daytona Beach, FL 32114, USA}

\author[0000-0003-1748-602X]{J. Farihi}
\altaffiliation{Visiting Astronomer at the Infrared Telescope Facility, which is operated by the University of Hawaii under contract 80HQTR19D0030 with the National Aeronautics and Space Administration.}
\affiliation{Department of Physics and Astronomy, University College London, London WC1E 6BT, UK}

\author{J.~L. Provencal}
\affiliation{Department of Physics and Astronomy and SARA, University of Delaware, Newark, DE 19716, USA}

\author{S. J. Kleinman}
\affiliation{Astromanager LLC, Hilo, HI, USA}

\author[0000-0002-1465-4780]{J. E. Pringle}
\affiliation{Institute of Astronomy, Cambridge, CB3 0HA, UK}

\author{A. Swan}
\altaffiliation{Visiting Astronomer at the Infrared Telescope Facility, which is operated by the University of Hawaii under contract 80HQTR19D0030 with the National Aeronautics and Space Administration.}
\affiliation{Department of Physics and Astronomy, University College London, London WC1E 6BT, UK}
\affiliation{Department of Physics, University of Warwick, Coventry CV4 7AL, UK}

\author{G. Fontaine}
\altaffiliation{Deceased.}
\affiliation{D\'epartement de Physique, Universit\'e de Montr\'eal, Montréal, QC Canada}

\author[0000-0001-5941-2286]{J. J. Hermes}
\affiliation{Department of Astronomy, Boston University, 725 Commonwealth Ave., Boston, MA 02215, USA}

\author{J. Sargent}
\affiliation{Department of Physics and Astronomy, The University of Georgia, Athens, GA 30602, USA}

\author{Z. Savery}
\affiliation{Department of Physics and Astronomy and SARA, University of Delaware, Newark, DE 19716, USA}

\author{W. Cooper}
\affiliation{Department of Physics and Astronomy and SARA, University of Delaware, Newark, DE 19716, USA}

\author[0000-0003-1202-9751]{V. Kim}
\affiliation{Fesenkov Astrophysical Institute, 50020 Almaty, Kazakhstan}

\author{V. Kozyreva}
\affiliation{Sternberg Astronomical Institute, Lomonosov Moscow State University, 119991 Moscow, Russia}

\author[0000-0002-2788-2176]{M. Krugov}
\affiliation{Fesenkov Astrophysical Institute, 50020 Almaty, Kazakhstan}

\author[0000-0002-7756-546X]{A. Kusakin}
\affiliation{Fesenkov Astrophysical Institute, 50020 Almaty, Kazakhstan}

\author[0000-0001-7143-0890]{A. Moss}
\affiliation{Department of Physics and Astronomy, The University of Oklahoma, Norman, OK 73019, USA}

\author[0000-0002-6293-9940]{W. Ogloza}
\affiliation{Mt. Suhora Observatory, Pedagogical University, PL-30-084 Krak\'ow, Poland}

\author[0000-0002-3326-2918]{Erika Pak\v stien\.e}
\affiliation{Institute of Theoretical Physics and Astronomy, Vilnius University, Saul\.etekio Av. 3, 10257 Vilnius, Lithuania}

\author[0000-0002-4313-7416]{A. Serebryanskiy}
\affiliation{Fesenkov Astrophysical Institute, 50020 Almaty, Kazakhstan}

\author{Eda Sonbas}
\affiliation{Astrophysics Application and Research Center and Department of Physics, Adiyaman University, 02040 Adiyaman, Turkey}

\author{B. Walter}
\affiliation{Meyer Observatory and Central Texas Astronomical Society, Waco, TX 76705, USA}

\author[0000-0001-5836-9503]{M. Zejmo}
\affiliation{Janusz Gil Institute of Astronomy, University of Zielona Gora, 65-516 Zielona Gora, Poland}

\author[0000-0003-3609-382X]{S. Zola}
\affiliation{Astronomical Observatory of the Jagiellonian University, ul.\ Orla 171, 30-244 Krak\'ow, Poland}





\begin{abstract}
More than 36 years have passed since the discovery of the infrared excess from circumstellar dust orbiting the white dwarf G29-38, which at 17.5\,pc it is the nearest and brightest of its class.  The precise morphology of the orbiting dust remains only marginally constrained by existing data, subject to model-dependent inferences, and thus fundamental questions of its dynamical origin and evolution persist.  This study presents a means to constrain the geometric distribution of the emitting dust using stellar pulsations measured at optical wavelengths as a variable illumination source of the dust, which re-radiates primarily in the infrared.  By combining optical photometry from the Whole Earth Telescope with 0.7--2.5\,$\upmu$m spectroscopy obtained with SpeX at NASA's Infrared Telescope Facility, we detect luminosity variations at all observed wavelengths, with variations at most wavelengths corresponding to the behavior of the pulsating stellar photosphere, but towards the longest wavelengths the light curves probe the corresponding time-variability of the circumstellar dust.  In addition to developing methodology, we find pulsation amplitudes decrease with increasing wavelength for principal pulsation modes, yet increase beyond $\approx$2\,$\upmu$m for  nonlinear combination frequencies.  We interpret these results as  combination modes deriving from principal modes of identical $\ell$ values and discuss the implications for the morphology of the warm dust.  We also draw attention to some discrepancies between our findings and theoretical expectations for the results of the non-linearity imposed by the surface convection zone on mode--mode interactions and on the behavior of the first harmonic of the highest-amplitude pulsation mode.
\end{abstract}



\section{Introduction}

White dwarf stars are known to host circumstellar dust that is widely interpreted as tidally disrupted, minor rocky bodies \citep{Jura14,Farihi16,Veras16,Guidry21}.  The standard scenario requires that one or more minor bodies are gravitationally perturbed by major planets \citep{Frewen14,Petrovich17,Smallwood18}, exciting eccentricities and eventually resulting in nearly radial orbits.  If a rocky body passes within $\approx$1\,R$_{\odot}$ of the white dwarf, it will be tidally shredded, leaving orbiting dust that then has a large cross section to self-collisions and Poynting-Robertson drag \citep{Bochkarev11,Veras15,Malamud20}.  

In the limit where the circumstellar debris is the result of just one disrupted body, this dust is expected to be in a plane and its orbit is assumed to eventually circularize, although this process is not yet fully understood \citep{Nixon20,Malamud21}.  If the dust is derived from multiple minor bodies originating on independent orbits, then there is no {\it a priori} reason why the circumstellar material should be confined to a single plane.  Therefore, constraining the dust morphology should constrain its origin, and thereby the dynamical evolution of the contributing bodies.  While the typical, $3-5$\,$\upmu$m detected dust emission implies orbital radii within or around the Roche limit of the star, currently there are only model-dependent inferences, and no compelling empirical constraints on the precise distribution and geometry of the circumstellar material \citep{Bonsor17,Swan20}.

Giclas\,29-38 (G29-38 = ZZ~Psc = WD~2326+049) is both a well-known luminosity variable \citep{Shulov74} of the ZZ\,Ceti class and the prototype white dwarf hosting circumstellar dust.  \citet{Zuckerman87} discovered its infrared excess, and subsequently \citet{Graham90} and \citet{Patterson91} compared pulsations observed in the $B$, $J$, $K$ (and $L$ in the latter case) photometric bands, both finding consistency with a relatively planar disk geometry.  

The idea behind using stellar pulsations in this manner, which forms the basis of our updated effort, is that geometric oscillations are seen differently by Earth-based observers than by the circumstellar dust, and that these two views may, in principle, be disentangled using multi-wavelength photometry.  Luminosity variations arising from the photosphere are prominent primarily in the optical bands where the stellar emission peaks \citep{Brassard95}, whereas any dust illuminated by the stellar pulsations will re-radiate effectively only at infrared wavelengths.   

The infrared instruments of these earlier observations recorded only weak pulsation signals, and conclusions regarding the pulsation modes were not possible, thus limiting the interpretation of the circumstellar dust morphology.  {\em Spitzer} infrared observations utilizing all three instruments, including photometry and spectroscopy, were used to characterize the overall infrared dust continuum, and identify broad 10\,$\upmu$m emission from small silicate dust grains (dominated by olivines), but ultimately the data are consistent with a range of possible circumstellar dust configurations \citep{Reach05,Reach09,Ballering22}.

Current constraints on the geometrical distribution of dust come purely from modeling the dust component of the spectral energy distribution \citep[e.g.][]{Jura03,Reach05,Ballering22}.  There are model-dependent inferences from luminosity and calcium equivalent width amplitudes for two large pulsations excited in 2008 that suggest polar accretion \citep{Thompson10}.  And while no stellar magnetic field has been detected toward G29-38 ($1\upsigma=0.5$\,kG; \citealt{Bagnulo21}), a field as modest as 0.01\,kG is more than sufficient to truncate a high-rate accretion flow of $10^9$\,g\,s$^{-1}$, and result in accretion along field lines \citep{Farihi18,Cunningham21}.  The stellar optical polarization has been sensitively measured at $275.3\pm31.9$\,ppm \citep{Cotton20}, and in the context of a dust disk model, is consistent either with a nearly face-on, optically thin disk or dust with a low Bond albedo.  While highly valuable, these studies do not offer further insight into circumstellar dust geometry, beyond the dust emission modeling from infrared observations.

The approach taken in this work is an updated version of the multi-wavelength pulsation studies \citep{Graham90,Patterson91}, and leverages the fact that infrared detectors and instrumentation have improved dramatically in the last four decades.  The primary instrument of this study, SpeX, saw first light in 2000 \citep{Rayner03} and its infrared array was upgraded in 2014.  In addition to technological advances in sensitivity and read-out compared to prior decades, a key advantage of SpeX is the synchronous multi-wavelength coverage possible with its low resolution spectroscopic mode.  Simultaneous observations of photosphere- and dust-emitted light obviates the need for phasing detector clocks and readouts.  Concurrent, multi-wavelength data are particularly valuable for G29-38, as its power spectrum is not stable and can change dramatically from one observing season to another \citep{Kleinman98}. These significant improvements in observing capability allow what is essentially a new technique for studying the circumstellar dust morphology of G29-38.  In this paper, we describe that technique and although we are unable to place firm constraints on the circumstellar dust morphology, we are able to connect the photospheric pulsations and dust responses to viable dust geometries.  In addition, we find that some pulsation properties appear to be discrepant with theoretical expectations as a function of wavelength, with relevance to all ZZ Ceti white dwarf stars.

\section{Observations and Reductions}
\label{obsres}

\subsection{SpeX Infrared Spectroscopy}

We observed G29-38 at the NASA Infrared Telescope Facility (IRTF), using the near-infrared imager and spectrograph SpeX \citep{Rayner03}, during two runs consisting of three contiguous partial nights each, on 2018 September 4--6, and again on 2020 November 8--10.  SpeX was operated in spectroscopic prism mode, simultaneously covering approximately 0.7 to 2.5\,$\upmu$m.  In 2018, the standard slit was opened to 3\,arcsec width ($R\approx20$), and in 2020 this was changed to 1.6\,arcsec ($R\approx40$).  The sky conditions for the 2018 observing run were variable with light cirrus on two nights, and seeing ranged from 0.7 to 1.1\,arcsec.  In 2020, the conditions were mostly excellent, with seeing between 0.3 and 0.8\,arcsec.  In both cases, the wide slits captured the majority of the stellar light, minimizing atmospheric dispersion losses that could alter the true spectral slope.

Integrations on the array were taken continuously during two or three observing blocks each night, preceded by, interspersed with, and immediately followed by calibration frames and standard star observations.  The science target was nodded along the slit in the standard ABBA pattern for point sources, for total durations of 45\,min to 3\,h, (typically 1--2\,h).  Integration times were 14.8\,s in 2018, and 5.10\,s in 2020, where the total time on-source was approximately 2.2 to 5.6\,h per night.  Calibration frames consisted of arc lamps for wavelength calibration, flat field images, and telluric standard stars.  A total of 1671 and 2200 spectra of G29-38 were obtained in 2018 and 2020, respectively.

The data were reduced with SpeXtool version v4.1 \citep{Vacca03,Cushing04}.  All reduction steps were carried out in the standard manner, including flat fielding, A-B pair subtraction, spectral extraction, and wavelength calibration.  The science target frames were all extracted individually, while those of the telluric standards were combined into a single spectrum of higher signal-to-noise (S/N).  The individual spectra of G29-38 were telluric corrected on a nightly basis, in a uniform manner using a single set of standard observations corresponding most closely to the median airmass of the science target.  This choice was motivated by the fact that time-varying signals would be introduced by using several telluric standards (at different airmasses) within a single night.  Instead, a single telluric standard was used to correct each nightly data set, and while this does not produce the best correction of telluric features for individual frames, it introduces a minimally time-variable (and well characterized) signal corresponding to the gradual change in airmass.  

The advantage of deriving frequency measurements from SpeX spectroscopy, obtained with a wide slit, is that such measurements are differential both in wavelength and time, and therefore insensitive to variable observing conditions such as seeing or transparency changes.  After the individual spectra were reduced, which corrects for non-linearities in the detector, removes night sky lines, etc., we created a simple procedure to interpolate across thorium radiation events and cosmic rays in the detector.  Only 51 spectra, or slightly over 1\%, were affected by such artifacts.  During this stage, spectra prior to MJD = 58365.48 (the beginning of 2018 Sep 04) were removed owing to clouds, together with some spectra from 2020 Nov 09 that were significantly fainter, likely due to guiding issues.  This procedure resulted in 1576 and 2184 spectra of G29-38 from 2018 and 2020, respectively.  

We defined ten bandpasses for synthetic photometry, along with two further bandpasses dominated by strong telluric lines, which were not analyzed further.  Table~1 lists the ten defined bandpasses, their central wavelength designation, and the actual S/N range in each of these resulting from synthetic photometry (cf.\ the bandpasses were designed to achieve S/N $\approx100$ per bandpass per observation).  The synthetic photometry S/N values are typically higher per observation for the 2018 spectra because these integrations were three times longer than for the 2020 spectra.  We discuss the subsequent normalization steps in the time-series analyses of these data below. 

\begin{table}
\caption{Bandpasses used in SpeX synthetic photometry.}
\begin{center}
\begin{tabular}{ccc}
\hline
Band   &\,$\uplambda$ & S/N \\ 
	   & ($\upmu$m)   & range \\ 
\hline
0.78   & 0.70--0.86   & 190--340 \\
0.94   & 0.86--1.02   & 200--360 \\
1.10   & 1.02--1.18   & 170--330 \\
1.26   & 1.18--1.34   & 150--280 \\
1.51   & 1.45--1.56   &  85--170 \\
1.62   & 1.56--1.68   &  85--170 \\
1.74   & 1.68--1.80   &  70--150 \\
2.06   & 1.97--2.14   &  75--140 \\
2.23   & 2.14--2.32   &  80--150 \\
2.41   & 2.32--2.50   &  45--75  \\
\hline        
\end{tabular}
\end{center}
\end{table}

\subsection{Moris Optical Photometry}

In order to extend the wavelength coverage of the IRTF observations, the SpeX observations used the Moris CCD imager for guiding purposes \citep{Gulbis11}.  The Moris images were taken through the SDSS $g$-band filter, over a $60\times60$\,arcsec$^2$ field of view on an $512\times512$\,pixel$^2$ CCD with 0.12\,arcsec pixels.  These data were obtained simultaneously with the SpeX data, but no effort was made to phase the readout times of these two instruments, as the guider was necessarily obtaining shorter exposures of 3.0\,s than employed with SpeX.  Due to their limited temporal coverage, these data offered little additional help in identifying the stellar pulsation modes, and we did not continue this approach during the 2020 IRTF run.

\subsection{WET Optical Photometry}

In both 2018 and 2020, we organized contemporaneous international, ground-based optical photometry campaigns to support the IRTF observations.  The telescopes, detectors, filters, and time-series durations are listed in Tables 2 and 3, where coverage totaled 112.2\,h in 2018, and 173.3\,h in 2020, and data reduction followed the prescription outlined in \citet{Provencal12}. In brief, raw images were calibrated and aperture photometry performed using the {\sc Maestro} photometry pipeline \citep{Dalessio10}, where each image was corrected for bias and thermal background, and normalized by its flat field. {\sc Maestro} performed photometry for a range of aperture sizes for the target and comparison stars. The combination of aperture and comparison star(s) that resulted in the highest quality raw light curve was chosen for each individual run from each observing site.

The next data reduction step employed the WQED pipeline \citep{wqed}. WQED examined individual light curves for photometric quality, removed outlying points, divided by appropriate comparison stars, and corrected for differential extinction.  The result was a series of light curves with times in seconds and amplitude variations represented as fractional intensity.  The final step in the reduction process combined the individual light curves from the WET photometry and applied barycentric corrections. For this step, it was assumed that G29-38 oscillates around a mean light level. This important assumption allowed the correction of instrumental intensity offsets for any overlapping light curves. The question of the treatment of overlapping data is discussed in detail in \citet{Provencal09}. The final product is a compiled optical light curve for G29-38 spanning the 2018 and 2020 SpeX observations.  

\begin{table}
\caption{Journal of 2018 time-series photometry.}
\begin{center}
\begin{tabular}{ccccc}
\hline
Date  & Telescope       & Detector & Filter & $\Delta$T \\
(UT)  &                 &          &        & (h) \\
\hline
08-02 & SARA-CT 0.6     & FLI      & $V$    & 4.7 \\
08-05 & SARA-RM 1.0     & Andor    & $V$    & 5.5 \\
08-09 & SARA-RM 1.0     & Andor    & $V$    & 6.4 \\
08-11 & SARA-RM 1.0     & Andor    & $V$    & 1.8 \\
08-17 & SARA-RM 1.0     & Andor    & $V$    & 9.2 \\
08-29 & TSAO 1.0        & Apogee   & $V$    & 2.6 \\
08-31 & TSAO 1.0        & Apogee   & $V$    & 1.4 \\
09-01 & Mol\.etai 0.35  & CCD4710  & $V$    & 4.7 \\
09-01 & TSAO 1.0        & Apogee   & $V$    & 2.4 \\
09-02 & Mol\.etai 0.35  & CCD4710  & $V$    & 3.2 \\
09-02 & TSAO 1.0        & Apogee   & $V$    & 3.8 \\
09-03 & Mol\.etai 0.35  & CCD4710  & $V$    & 4.0 \\
09-03 & TSAO 1.0        & Apogee   & $V$    & 5.1 \\
09-04 & IRTF 3.2        & MORIS    & $g$    & 2.5 \\
09-04 & TSAO 1.0        & Apogee   & $V$    & 2.7 \\
09-05 & IRTF 3.2        & MORIS    & $g$    & 5.4 \\
09-05 & TSAO 1.0        & Apogee   & $V$    & 4.5 \\
09-06 & IRTF 3.2        & MORIS    & $g$    & 2.8 \\
09-06 & TSAO 1.0        & Apogee   & $V$    & 3.5 \\
09-07 & Mol\.etai 1.65  & CCD4710  & $V$    & 5.7 \\
09-07 & TSAO 1.0        & Apogee   & $V$    & 3.5 \\
09-10 & Krakow 0.5      & Alta U42 & $V$    & 4.8 \\
09-13 & Krakow 0.5      & Alta U42 & $V$    & 3.0 \\
09-16 & Krakow 0.5      & Alta U42 & $V$    & 6.1 \\
09-30 & TSAO 1.0        & Apogee   & $V$    & 3.4 \\
10-05 & ERAU 1.0        & STX16803 & $i$    & 2.7 \\
12-05 & SARA-RM 1.0     & Andor    & $V$    & 3.4 \\
12-10 & SARA-RM 1.0     & Andor    & $V$    & 3.4 \\
\hline
\end{tabular}
\end{center}
{\em Notes:} The ERAU 1.0-meter is at Embry Riddle Aeronautical University in the USA. The Krakow 0.5-meter is at the Astronomical Observatory of the Jagiellonian University in Poland. The Mol\.etai 0.35-meter and 1.65-meter are at the Mol\.etai Astronomical Observatory in Lithuania. The SARA-CT 0.6-meter is at the Cerro Tololo Inter-American Observatory in Chile. The SARA-RM 1.0-meter is at the Roque de los Muchachos Observatory in Spain.  The TSAO 1.0-meter is at the Tien Shan Astronomical Observatory in Kazakhstan.
\end{table}

\begin{table}
\caption{Journal of 2020 time-series photometry.}
\begin{center}
\begin{tabular}{ccccc}
\hline
Date  & Telescope     & Detector      & Filter & $\Delta$T \\
(UT)  &               &               &        & (h) \\
\hline
11-02 & ADYU 0.6      & Andor Tech    & $g$    & 1.53, 1.6 \\
11-03 & Prompt P6 0.4 & STX-16803     & $V$    & 1.6 \\
11-04 & MCAO 0.6      & Aspen         & BG40   & 1.1 \\
11-04 & Prompt P6 0.4 & STX-16803     & Lum    & 1.8, 2.5 \\
11-04 & Prompt P6 0.4 & STX-16803     & $V$    & 0.3 \\
11-05 & AZT-20 1.5    & FLI           & $g$    & 3.1 \\
11-05 & Krakow 0.5    & Alta F-42     & BG40   & 2.4 \\
11-05 & Krakow 0.5    & Alta F-42     & $V$    & 0.6, 3.4 \\
11-05 & MCAO 0.6      & Aspen         & BG40   & 1.3 \\
11-05 & Prompt P6 0.4 & STX-16803     & Lum    & 0.8 \\
11-05 & Suhora 0.4    & Moravian G2   & $R$    & 1.2 \\
11-05 & Suhora 0.6    & Aspen-47      & BG40   & 3.9 \\
11-06 & AZT-20 1.5    & FLI           & $g$    & 6.1 \\
11-06 & Krakow 0.5    & Alta F-42     & BG40   & 1.4 \\
11-06 & PJMO 0.6      & Pixis 2048BeX & BG40   & 5.0 \\
11-06 & Prompt P5 0.4 & STX-16803     & Lum    & 0.5 \\
11-06 & Suhora 0.4    & Moravian G2   & $R$    & 2.3 \\
11-06 & Suhora 0.6    & Aspen-47      & BG40   & 3.5 \\
11-07 & ADYU 0.6      & Andor Tech    & $g$    & 3.6 \\
11-07 & AZT-20 1.5    & FLI           & $g$    & 3.2 \\
11-07 & Krakow 0.5    & Alta F-42     & BG40   & 4.3 \\
11-07 & Krakow 0.5    & Alta F-42     & $V$    & 2.2 \\
11-07 & Prompt P5 0.4  & STX-16803     & Lum    & 2.3, 3.6 \\
11-07 & Suhora 0.4    & Moravian G2   & $R$    & 6.2 \\
11-07 & Suhora 0.6    & Aspen-47      & BG40   & 7.2 \\
11-08 & ADYU 0.6      & Andor Tech    & $g$    & 2.2 \\
11-08 & Krakow 0.5    & Alta F-42     & $V$    & 4.5 \\
11-08 & PJMO 0.6      & Pixis 2048BeX & BG40   & 5.9 \\
11-08 & Prompt P2 0.4 & STX-16803     & Lum    & 0.5, 2.8, 4.3 \\
11-08 & Suhora 0.6    & Aspen-47      & BG40   & 8.0 \\
11-09 & ADYU 0.6      & Andor Tech    & $g$    & 5.1 \\
11-09 & PJMO 0.6      & Pixis 2048BeX & BG40   & 6.6 \\
11-09 & Suhora 0.4    & Moravian G2   & $R$    & 5.4 \\
11-09 & Suhora 0.6    & Aspen-47      & BG40   & 9.2 \\
11-10 & PJMO 0.6      & Pixis 2048BeX & BG40   & 5.5 \\
11-10 & Suhora 0.4    & Moravian G2   & $R$    & 6.7 \\
11-10 & Suhora 0.6    & Aspen-47      & BG40   & 7.9 \\
11-11 & ADYU 0.6      & Andor Tech    & $g$    & 7.5 \\
11-13 & MCDO 2.1      & Roper E2V     & BG40   & 1.8 \\
11-20 & SARA-RM 1.0   & Andor         & $V$    & 2.4 \\
11-21 & SARA-RM 1.0   & Andor         & $V$    & 3.4 \\
\hline
\end{tabular}
\end{center}
{\em Notes:} The ADYU 0.6-meter is at the Adiyaman Observatory, Adiyaman Turkey. The AZT-20 1.5-meter is at the Assy-Turgen Observatory is in Kazakhstan. The MCAO 0.6-meter is at the Mt. Cuba Astronomical Observatory, USA. The MCDO 2.1-meter is at McDonald Observatory, USA. The PJMO 0.6-meter is at the Central Texas Astronomical Society's Meyer Observatory in the USA. The Prompt P2, P5, and P6 0.4-meter telescopes are at the Cerro Tololo Inter-American Observatory in Chile. The Suhora 0.35-meter and 0.6-meter telescope are at Suhora Observatory, Poland. 
\end{table}

\section{Analysis}

\subsection{Frequency Identification}

{\sc Period04} \citep{Lenz05} and {\sc Pyriod} \citep{Bell2021} were used to derive discrete Fourier transforms (FTs) of the time series photometry, with amplitude units of milli-modulation amplitude (1\,mma = 0.1\%).  Figure~1 presents the FTs of the 2018 WET photometry (upper left and lower right panels) as well as FTs of the 2018 SpeX synthetic phototometry with wavelength in microns  indicated in each panel.  For the synthetic photometry at 1.51\,$\upmu$m, data from the first night of this 2018 run were not included because of excess noise at this wavelength, presumably due to excess sky noise in the adjacent sky band. The pulsation frequencies at 1113 and 1211\,$\upmu$Hz are clearly visible in the optical and near-infrared bandpasses though weaken significantly at $\geq 1.6 \upmu$m.  The pulsation frequencies at 788 and 2225\,$\upmu$Hz display the opposite behavior and appear strongest at the longest wavelengths.  Figure~2 presents similar FTs for the 2020 data.  At this epoch, all pulsation frequencies were weaker than in 2018, though the peak near 2000\,$\upmu$Hz is clearly visible.  The amplitude of this frequency decreases as a function of wavelength out to 1.6\,$\upmu$m, then it appears to increase.  The lower pulsation amplitude of G29-38 is within its previously observed range \citep{Kleinman98}.

The 2018 and 2020 optical light curves form the basis for the identification of the pulsation frequencies during the SpeX observations. A statistically significant frequency is defined as a frequency with an amplitude at least four times the average noise level, representing a 99.9\% probability that the frequency represents a true signal in the data and is not the result of random noise.  The noise level is defined as the the average amplitude remaining in the FT after pre-whitening (discussed further below) by the dominant peaks \citep{Provencal12}, and is a conservative estimate, especially for the 2018 data.  In 2018, G29-38 was in a fairly high amplitude pulsation state and exhibited amplitude modulation, and thus it was impossible to completely remove all of the significant power before calculating the noise level.  The mean noise over a broad range of pulsation frequencies in 11 different frequency ranges (WET optical plus ten SpeX bands) for 2018 and 2020 are presented in Figure~3.  During the 2018 campaign, the WET observations had a typical mean noise level of $\la1$\,mma.  The IRTF observations, covering fewer pulsation cycles, had somewhat higher noise levels, rising up to $\approx$4\,mma in a few cases.  In particular, for the bluest SpeX band, at 0.78\,$\upmu$m, there was additional noise near 1000\,$\upmu$Hz, and for the reddest two SpeX bands, at 2.23 and 2.41\,$\upmu$m, there was additional noise near 2300\,$\upmu$Hz.  The likely cause of this additional noise at these isolated frequencies is our inability to completely remove the signal produced by amplitude modulation.

\begin{figure}  
\includegraphics[width=\linewidth]{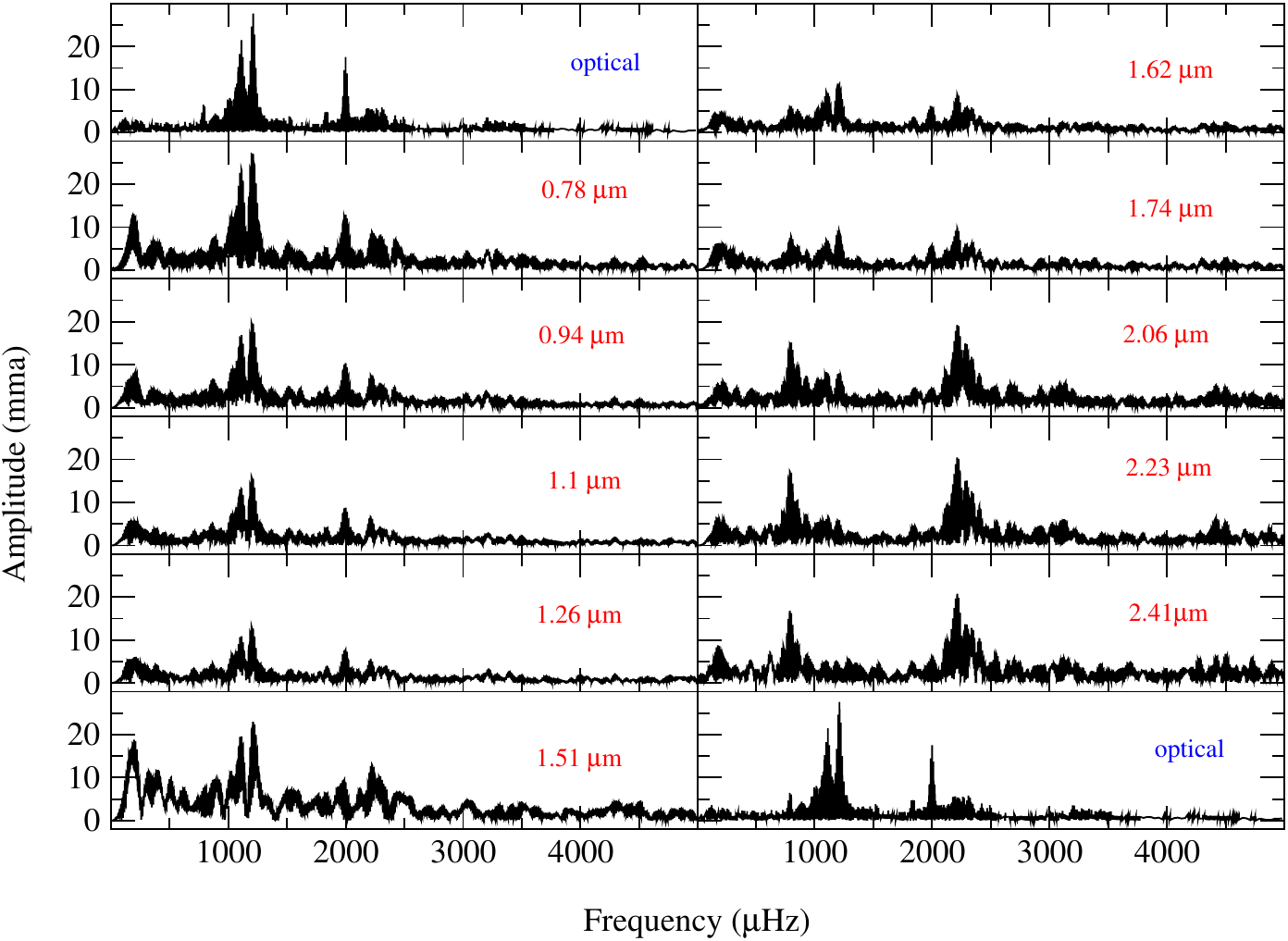}
\caption{FTs for the 2018 WET photometry, labeled "optical", and the ten 2018 SpeX bands (labeled with their central wavelengths).  The WET optical FT is repeated in the lower right panel to aid in visualizing the amplitudes of the pulsation frequencies versus wavelength.  The pulsation frequencies at 1113 and 1211\,$\upmu$Hz are clearly visible in the optical and near-infrared bandpasses, though weaken at wavelengths $\geq1.6\,\upmu$m.  The pulsation frequencies at 788 and 2225\,$\upmu$Hz display the opposite behavior and appear strongest at the longest wavelengths. 
\label{fig1}}
\end{figure}

\begin{figure}  
\includegraphics[width=\linewidth]{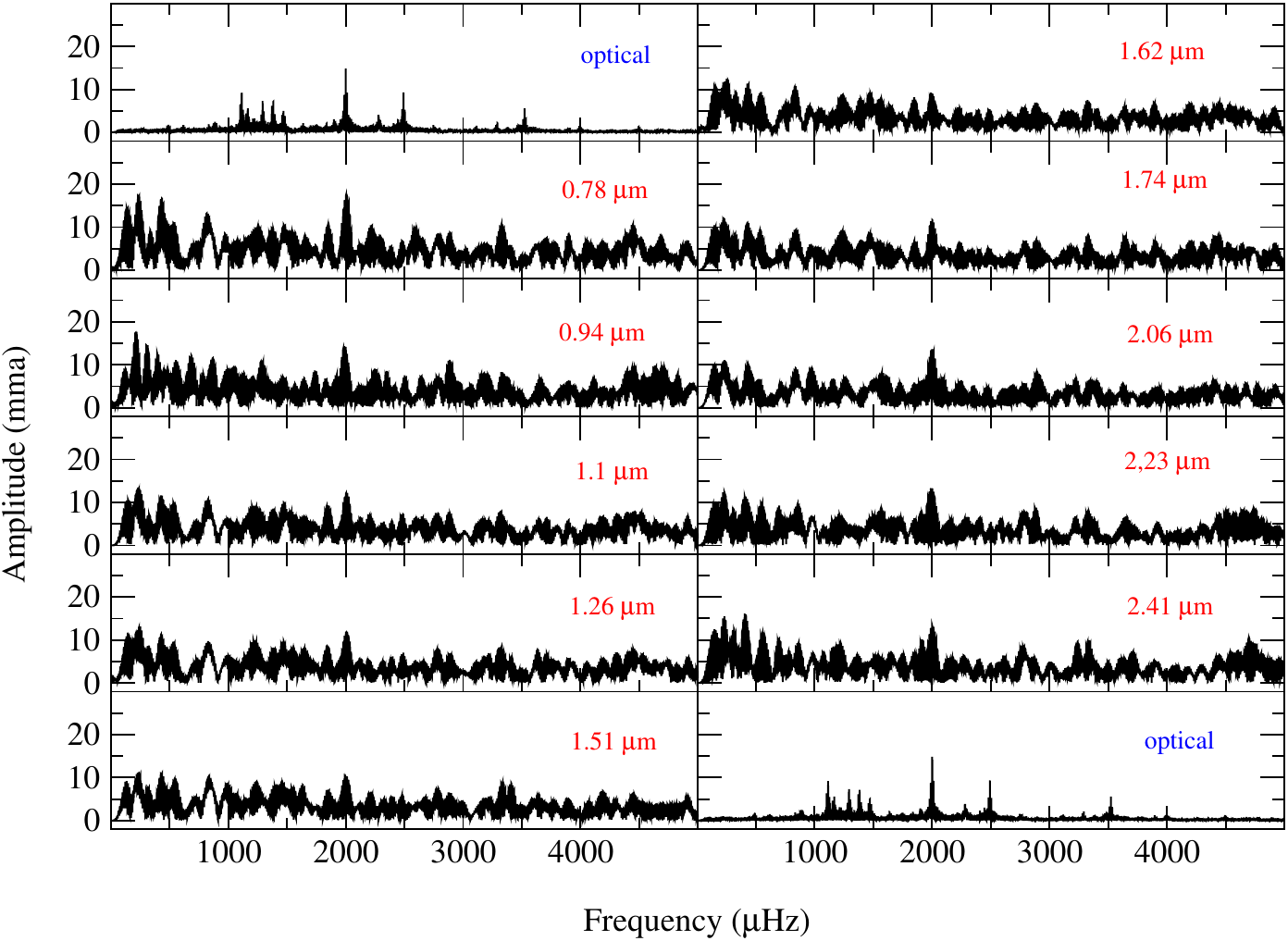}
\caption{Same as Figure~1, except based on the November 2020 data.  The strongest pulsation frequencies are approximately half the amplitudes of those in September 2018.  
\label{fig2}}
\end{figure}

\begin{figure}  
\includegraphics[width=\linewidth]{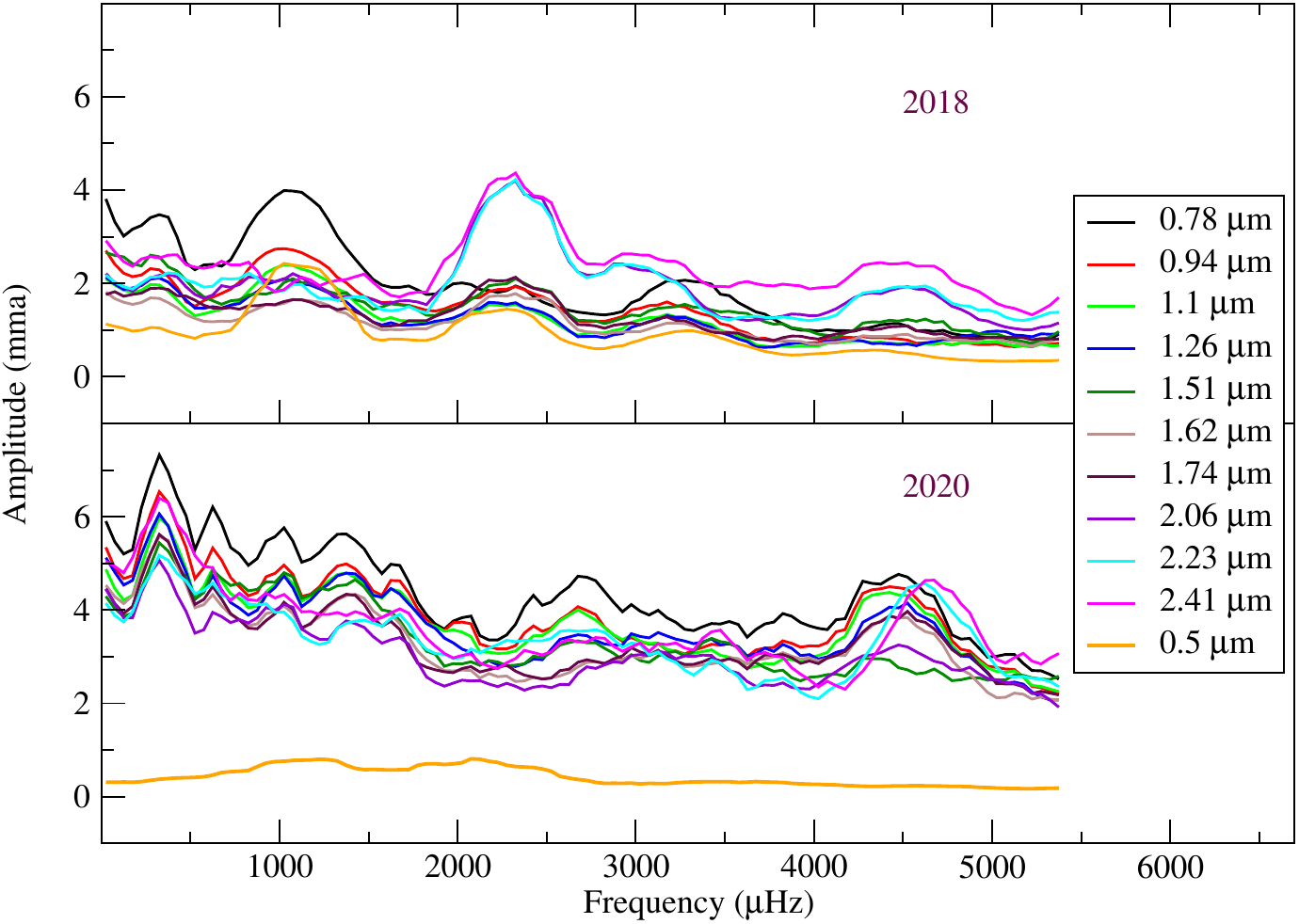}
\caption{Mean noise as a function of frequency for the optical photometry and the SpeX synthetic photometry. The increased noise in several of the SpeX channels, particularly in 2018 at 0.78\,$\upmu$m near 1000\,$\upmu$Hz and at 2.23 and 2.41\,$\upmu$m near 2300\,$\upmu$Hz, are likely due to our inability to completely remove the dominant frequencies.  
\label{fig3}}
\end{figure} 

We also calculated Monte Carlo simulations for the WET data using the Monte Carlo routine in {\sc Period04}. This routine generates a series of simulated light curves using the original times, the fitted frequencies and amplitudes and also containing Gaussian noise.  Each simulated light curve is subjected to a least-squares fit. The uncertainties are produced by the distribution of fit parameters.  The Monte Carlo results are consistent with a noise level of 0.5\,mma for the 2018 optical data and 0.2\,mma for the 2020 optical data. 

Armed with an understanding of the noise properties of these data, frequency identification is now possible. In both data sets, the noise in the FTs derived from the WET runs is significantly lower than the FT noise from the SpeX synthetic photometry, which covered fewer stellar oscillations. Given the additional higher frequency resolution of the optical data, we use the WET optical frequencies as the basis to identify the frequencies in the SpeX FTs.  

{\sc Period04} and {\sc Pyriod} were used to pre-whiten the FTs.  The process involves identifying the largest amplitude resolved peak in the FT, or the Lomb-Scargle periodogram in the case of {\sc Pyriod}, fitting the data set with a sinusoid of that frequency, subtracting the fit from the light curve, recomputing the FT, examining the residuals, and repeating the process until no significant power remained. This process works well for stable~pulsators, but is more complicated in the presence of amplitude modulation.  We illustrate the process in Figure~4 with an example of pre-whitening of the dominant power in the 2018 WET data. In the top panel of Figure~4, the red arrow identifies the dominant 1211\,$\upmu$Hz peak.  The middle panel shows the result of removing a sinusoid with that frequency, and identifies the next highest amplitude frequency at 1112\,$\upmu$Hz.  The bottom panel shows the result of removing both the 1211\,$\upmu$Hz and 1112\,$\upmu$Hz peaks from the data. Significant power remains, but a resolved frequency cannot be clearly identified.  This is the signature of amplitude modulation, and the remaining power cannot be pre-whitened.  

Table~4 contains the resulting frequency identifications,  within the constraints imposed by amplitude modulation, for the WET 2018 and 2020 data sets. The table~presents frequency and its uncertainty in\,$\upmu$Hz, amplitude and its uncertainty in mma, and the S/N of the measured amplitude, which is the ratio of the amplitude of the frequency to the amplitude of the noise in that frequency range. We use the symbols $f$ and $g$ to identify frequencies in the 2018 and 2020 data sets, respectively.

\begin{figure}  
\includegraphics[width=\linewidth]{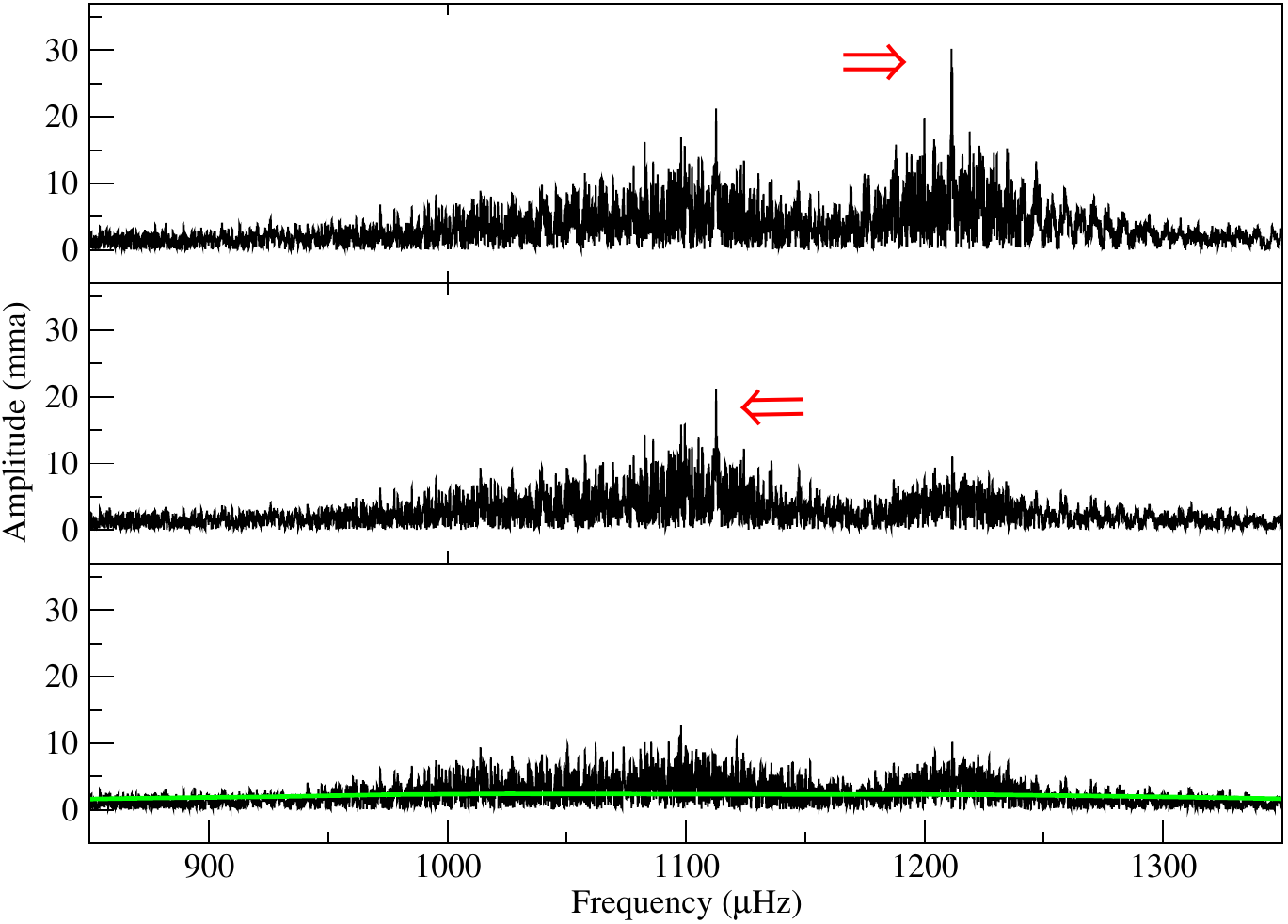}
\caption{Demonstration of pre-whitening using the dominant 1211\,$\upmu$Hz frequency in the 2018 WET data.  The first panel is the original FT.  The red arrow identifies the dominant 1211\,$\upmu$Hz peak. The second panel has been pre-whitened by 1211\,$\upmu$Hz, and the third panel is pre-whitened by both 1211\,$\upmu$Hz and 1112\,$\upmu$Hz. The green line indicates the formal noise level.
\label{fig_pre-white}}
\end{figure} 

\begin{table}
\caption{G29-38 frequency analyses from optical time-series photometry.}
\begin{center}
\begin{tabular}{cccc}
\hline
ID  & Frequency       & Amplitude    & S/N \\
    & ($\upmu$Hz)       &  (mma)       &     \\
\hline
{\bf2018} \\
$f_0$      & 1211.395 (0.005) & 32.7  (0.5)  & 64 \\
$f_1$      & 1112.656 (0.005) & 29.3  (0.5)  & 38 \\
$f_2$      & 1999.522 (0.008) & 17.8  (0.5)  & 35 \\
$f_3$      & 1013.340 (0.015) &  8.7  (0.5)  & 16 \\
$f_4$      & 1835.73  (0.03)  &  5.9  (0.5)  & 10 \\
$f_5$      & 1519.09  (0.04)  &  4.35 (0.5)  & 8  \\
$f_6$      & 2492.05  (0.03)  &  5.7  (0.5)  & 10 \\
$2f_0$     & 2422.735 (0.009) &  4.5  (0.5)  & 9  \\
$f_0+f_1$   & 2324.051 (0.007) &  8.1  (0.5)  & 16 \\
$f_0+f_3$   & 2224.735 (0.015) &  6.5  (0.5)  & 11 \\
$f_2-f_0$ & 788.127  (0.009) &  7.6  (0.5)  & 14 \\
{\bf2020} \\
$g_0$      & 2001.02  (0.01)  & 15.1  (0.3)  & 50\\
$g_1$      & 2492.41  (0.01)  &  9.3  (0.3)  & 31 \\
$g_2$      & 1114.60  (0.01)  &  9.1  (0.3)  & 29 \\
$g_3$      & 1838.06  (0.01)  &  7.3  (0.3)  & 24 \\
$g_4$      & 1292.21  (0.1)   &  7.4  (0.3)  & 24 \\
$g_5$      & 1296.67  (0.2)   &  4.0  (0.3)  & 13 \\
$g_6$      & 1166.51  (0.2)   &  5.8  (0.3)  & 19 \\
$g_7$      & 1164.05  (0.02)  &  4.4  (0.3)  & 14 \\
$g_8$      & 3522.78  (0.02)  &  5.6  (0.3)  & 17 \\
$g_9$      & 1470.77  (0.02)  &  5.1  (0.3)  & 16 \\
\hline
\end{tabular}
\label{tab:freqtable}
\end{center}
{\em Notes:} The identified frequencies are labeled sequentially in amplitude from $f_0$ for 2018 and from $g_0$ for 2020.
\end{table}

The next step was to fit the pulsation amplitudes and phases for the ten bands of synthetic SpeX photometry in order to determine how these varied as a function of wavelength.  This was performed independently for the 2018 and 2020 data sets because G29-38 is, as stated above, not a stable pulsator on a yearly timescale.  We fixed the frequencies to those derived from the WET time-series photometry because of the much greater timebase of these data, and then used {\sc period04} to fit the amplitude and phases in the SpeX data. 

\subsection{Patterns Among Pulsation Properties}

Figure~1 displays a surprising result.  The frequencies at 1211, 1112, and 1999\,$\upmu$Hz ($f_0$, $f_1$, and $f_2$ in Table~4) are clearly visible in the optical and near-infrared bandpasses and weaken significantly at $\geq 1.6 \upmu$m.  However, the frequencies at 788 and 2225\,$\upmu$Hz display the opposite behavior. These two frequencies have been identified as combination modes, defined as exact numerical combinations (by addition or subtraction) of larger amplitude, principal modes. In pulsating white dwarf stars, these combination modes are thought to arise from mode--mode interactions in the convection zone \citep{Brickhill92b, Wu01}. The two combination frequencies at 788 and 2225\,$\upmu$Hz (corresponding to $f_2-f_0$ and $f_0+f_3$ in Table 4) increase in amplitude as a function of wavelength. 

Figure~2 presents similar FTs for the 2020 data.  At this epoch, G29-38 was in a lower amplitude pulsation state.  These lower pulsation amplitudes provided S/N $\leq$ 10 for the amplitude measurements of the $g_0$ (2001\,$\upmu$Hz) mode in the Spex data, with weaker modes even harder to measure.  For this reason, we drop further analysis of the 2020 data and focus on the amplitude and phase behavior of the pulsations in the 2018 data.

Figure~5 presents the amplitudes (top panels) and times of amplitude maxima (bottom panels) as a function of wavelength for pulsation frequencies with sufficient S/N to discern patterns among their properties.  The left-hand panels display properties for the three highest amplitude g-modes (principal modes f$_0$, f$_1$, and f$_2$) and the right-hand panels display the four other harmonic or nonlinear combination frequencies (2f$_0$, f$_0$+f$_1$, f$_0$+f$_3$, and f$_2$$-$f$_0$).  For the principal modes, amplitudes decrease with increasing wavelength, as expected for photospheric g-modes (see \citealt{Brassard95}, figure 3) with no significant luminosity contribution via dust absorption and re-radiation.  
Additionally, the time of maxima for f$_0$, f$_1$, and f$_2$ are approximately constant versus wavelength except for the optical WET measurement for f$_0$ which lags the time of maxima of f$_0$ (period 825 s) at most other wavelengths by about one minute.  This one phase offset could be caused by a change in the orientation of the f$_0$ mode captured during the 2018 WET run, which had a longer temporal coverage than the 2018 IRTF observations.  Except for this one WET time lag, the three principal modes again behave as one would expect for atmospheric pulsations with little to no contribution by re-radiated light from the dust.

The amplitudes of the combination frequencies (f$_0$+f$_1$, f$_0$+f$_3$, f$_2$$-$f$_0$), follow a different pattern.  The observed amplitudes of these combination modes generally decrease from the visible through near-infrared, then increase, particularly beyond 2\,$\upmu$m.  While the increase in mode amplitudes at the longest wavelengths are what one would expect for modes strongly re-radiated by the dust, the spectral energy distribution modeling of \citet{Reach09} estimated that 34\% of the flux at 2.2\,$\upmu$m is due to the dust, with the remaining flux emitted by the star's  photosphere.  Assuming these pulsation amplitudes have the same fraction of dust contribution as the time-averaged star, even after correcting for the dust contribution the photospheric pulsation amplitudes increase through 2.5\,$\upmu$m.  Although tentative, we note that this is in contrast to the prediction of \citet{Wu01}, who finds that the ratios of the amplitudes of parent modes to the amplitudes of their resultant combination modes are independent of wavelength.  
The 2f$_0$ harmonic (discussed further below) may also increase in the infrared or may be consistent with a near constant amplitude at all observed wavelengths.  The time of maxima for these frequencies varies considerably, from essentially constant (f$_0$+f$_3$) to decreasing as a function of wavelength (2f$_0$, f$_0$+f$_1$).  One combination mode in the fourth panel, f$_2$$-$f$_0$, has had 1100 sec subtracted from its times of maxima to aid visualization.  

\begin{figure} 
\includegraphics[width=\linewidth]{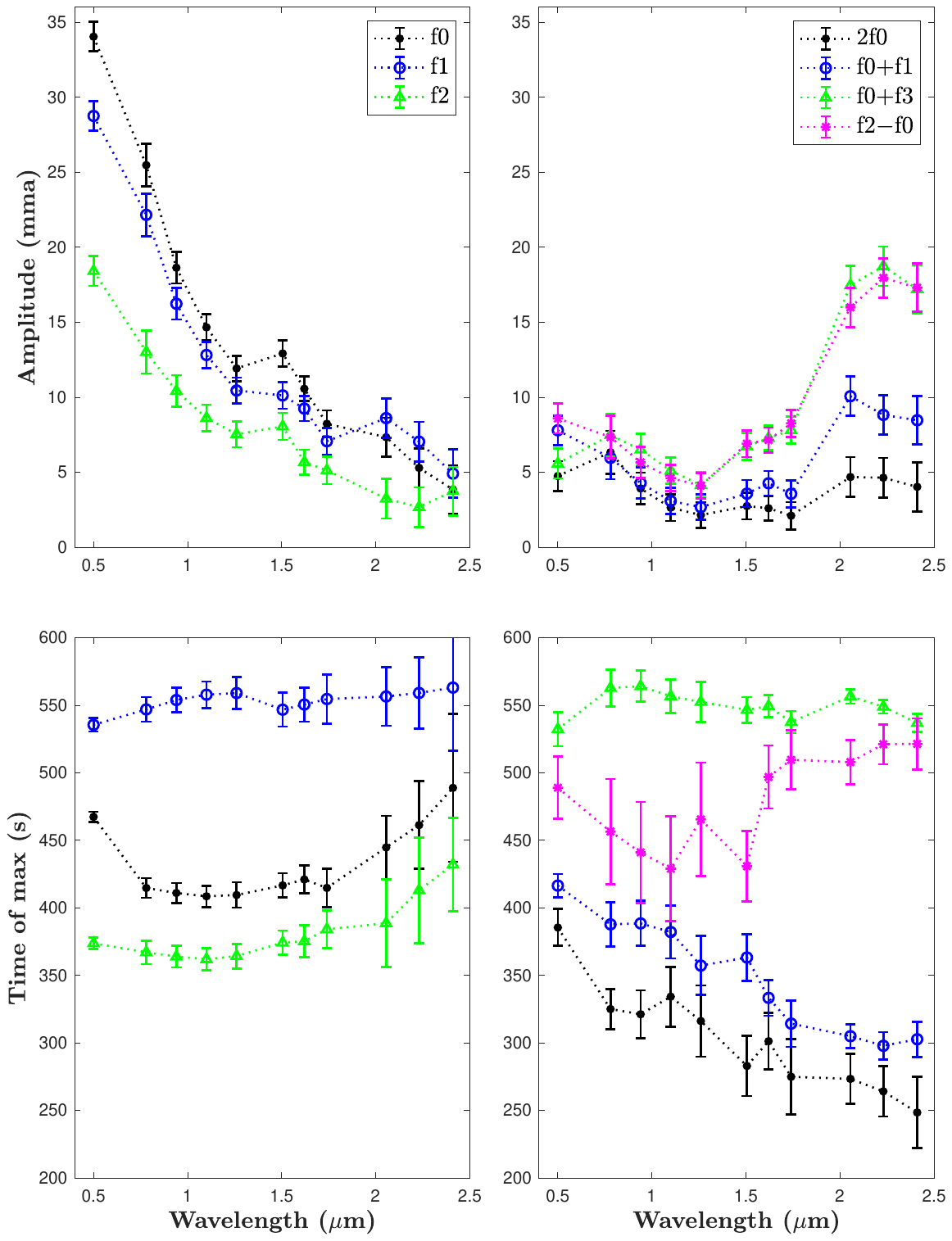}
\caption{The top two panels display pulsation amplitudes measured during 2018 as a function of wavelength.  The three highest amplitude g-modes (left-hand panels) clearly decrease in amplitude with wavelength.  The amplitude of the first harmonic (2f$_0$ in the right hand panels) of the highest amplitude g-mode (f$_0$) displays a different behavior.  The amplitudes of the combination modes (f$_0$+f$_1$, f$_0$+f$_3$ and f$_2$-f$_0$ in the right hand panels) increase beyond 2\,$\upmu$m.  The bottom panels display the times of maxima for these pulsations as a function of wavelength. 
\label{fig5}}
\end{figure}

\section{Discussion}

We now examine whether the optical and near-infrared mode characteristics (amplitude and phase) observed in 2018 are broadly consistent with a few simple models.  We do not make this comparison to the 2020 data because the star's pulsations were too weak at the observed epoch.  We seek to find a simple model that can explain how the three highest amplitude pulsation frequencies (f$_0$, f$_1$, f$_2$), which are all principal pulsation modes, decrease in amplitude as a function of increasing wavelength, while the combination modes display an increasing amplitude with wavelength, particularly beyond 2\,$\upmu$m (Figure~5).  Making the interpretation more difficult is that the mode identifications (the specific spherical harmonics) for any of the observed modes are not well known.  The general dearth of consistent multiplets in G29-38, along with its empirically variable frequencies and amplitudes with time, have made mode identification challenging \citep{Kleinman98}. Recent work using long-baseline Transiting Exoplanet Survey Satellite data has revealed some possible mode identification \citep{Uzundag23}, but since those observations are separated from our IRTF data by many months (and G29-38 has shown changes on shorter timescales), we have not adopted those identifications here.

\subsection{Information Encoded in the Combination Modes}

The combination modes (all combinations with f$_0$) show the strongest infrared signatures.  In general, combination modes have different geometries (in terms of the angular dependence of their surface brightness variations) than their underlying parent modes.  Currently, with insufficient knowledge of the mode $\ell$ identifications, it is not possible to uniquely model the stellar surface geometry and potentially constrain the dust geometry.  We search instead for a simple picture that unites these infrared signals from the combination models.

Principal pulsation modes have surface variations that can be described via spherical harmonic functions, $Y_l^m(\Theta,\Phi$), where $\Theta$ and $\Phi$ are angles of latitude and longitude with respect to some chosen axes in the star, $\ell$ is the spherical harmonic order, or total number of surface nodal lines, and $m$  identifies the number of nodes along a line of latitude.  Note that for stellar g-modes $\ell$=1,2,3,... and $|m| \leq \ell$ for each mode.  For a given mode with order $\ell$, the relative amplitudes of the components with values $m$ depend on the axes chosen.  If the axis of rotation is known, it is commonly used to define the $\Theta$, $\Phi$ coordinate system.  Alternatively, if circumstellar dust were primarily in a plane, that and a perpendicular axis could define longitude and latitude for mathematical convenience.  There is, as yet, no such known natural coordinate system for G29-38.  

The angular dependence of a combination mode with individual dependencies $Y_{\ell_1}^{m_1}$ and $Y_{\ell_2}^{m_2}$ is the product of these two functions \citep{Brickhill92b, Brassard95, Wu01}, which equals the sum over a set of $Y_\ell^m$ with $\ell$ running from $|\ell_1-\ell_2|$ to $\ell_1+\ell_2$.  While the resulting geometry can be complicated, if $\ell_1 = \ell_2$, then there is always a component in the sum which has $\ell = 0$.  (See also \citealt{Brassard95}, appendix C where they demonstrate that either a product of two $\ell = 1$ modes, one $m = +1$, the other $m = -1$, or a product of two $\ell = 2, m = 0$ modes have a spherically symmetric $Y_0^0$ component, but the product of an $\ell = 1$ mode and an $\ell = 2$ mode has no such component.)  Therefore, if the combination mode results from two modes with the same $\ell$ value, there will be a component of the combination mode that has {\it no angular dependence} (see also \citealt{Kurtz05}).  This component is an isotropic pulsation, much like a radial-mode pulsator such as an RR\,Lyrae or Cepheid.  In such a case, whatever the dust distribution, such a component will always provide a signal in the re-radiated flux.  We propose the possibility that the observed 2018 combination modes were visible in the re-radiated flux precisely because each contributing mode had the same value for $\ell$.  This suggestion offers a prediction and possibly a constraint for future mode identification.

\subsection{Historic Data}

Historic time-series optical and infrared data were presented by \citet{Graham90} and \citet{Patterson91}. The data presented by \citet{Patterson91} seems to be slightly higher quality, judging by the number of pulsation frequencies detected, but is in agreement with \citet{Graham90}.  \citet{Patterson91} detected pulsations in $B$ with periods of 615\,s (1620\,$\upmu$Hz), 186, 243, and 268\,s. They also detected pulsations in $K$ with periods of 186, 242, and 272\,s, which are consistent with the pulsations periods seen in $B$. They interpreted the 615\,s period as a principal stellar mode, with no corresponding dust response. They envisaged the dust as being in the form of an optically thin circular disk (cf.\ Section 4.4.1), and suggested that this mode (with respect to axes aligned with the disk) had $m \neq 0$, so that it elicited no response from the dust disk. They argued that the amplitudes of the other  pulsation frequencies were too high at $K$ to be just stellar g-modes (indeed \citealt{Graham90} found these pulsations only in $K$ and not in $B$), and thus that they must represent a dust response. However, since the theory of combination modes, and in particular their surface structures, was only just being developed (\citealt{Brickhill92b}; and later \citealt{Brassard95, Goldreich99, Wu01}), they assumed the stellar surface brightness variations to be of the form relevant to the principal modes, viz., $Y_\ell^m (\Theta, \Phi)$. Thus in order to explain why the pulsations seen in $K$ were not seen in $B$ they needed to appeal to a special geometry for the assumed stellar g-modes relative to the disk and to the observer. In particular, the infrared-strong modes all needed to have $\ell=2, m=0$ in axes relative to the disk, and the disk needed to be oriented at an angle of $\approx$55 degrees relative to the observer.
 
In our interpretation, while the 615\,s period arises from a principal stellar mode (most likely with $\ell=1$, \citealt{Kleinman98}), the other detected pulsation frequencies that are all strong in the infrared would be combination modes, whose fundamental contributors all have the same value of $\ell$. In this interpretation a special geometric configuration is not required.

\subsection{The First Harmonic}

The oscillation seen at frequency 2f$_0$ is the first harmonic of the parent mode f$_0$, which gives rise to the combination modes that we find to have strong visibility in the near-infrared. 
 
G29-38 is well-known as a high amplitude oscillator, and as such it would be no surprise if the dominant mode that we find, the f$_0$ mode, were of sufficiently high-amplitude that its underlying variability is slightly non-linear. Such non-linearity would primarily manifest itself as a first harmonic in the light curve with period 2f$_0$. If the f$_0$ mode has surface angular dependence $Y_\ell^m$, then this contribution to the variability at frequency 2f$_0$ would also have the same $Y_\ell^m$ surface distribution.
 
As an additional effect, for a single principal mode \citep{Wu01}, the sinusoidal oscillation at the base of the convection zone is modified as it passes through the zone. Thus the emergent oscillation is no longer purely  sinusoidal, but can be represented as a Fourier series with harmonic frequencies 2f$_0$, 3f$_0$, . . . . These are seen because the original oscillation has been distorted. The magnitude of the coefficient for the first harmonic 2f$_0$ contribution caused by this effect is proportional to the square of the amplitude of the original mode, and is therefore non-negative. This implies that, if the original mode f$_0$ has a surface distribution of the form $Y_\ell^m$, the first harmonic contribution 2f$_0$ caused by this effect has surface distribution $Y_\ell^m \times Y_\ell^m$. Thus this contribution to the first harmonic always has a surface component that is isotropic. For this reason the first harmonic frequency 2f$_0$ can, in principle, be more prominent than the underlying parent mode. As \cite{Wu01} comments: “This arises because the apparent amplitudes of the higher $\ell$ modes suffer stronger cancellation when integrated over the stellar disc, while the  harmonics of these modes do not”. In this respect the first harmonic induced by passage through the convection zone is analogous to the behavior of combination modes with the same value of $\ell$ (Section 4.1). We note that in our data the near-infrared response elicited by the 2f$_0$ harmonic is significantly less than for the combination modes that we detect (Section 3.2). It is possible that this comes about because of the two possible sources of the oscillation at frequency 2f$_0$, which, as we have noted, have two different surface brightness distributions.

The wavelength behavior of 2f$_0$ may aid future modeling of g-mode stellar pulsations in constraining properties of the convection zone \citep{Montgomery20}, and in that spirit we present additional analysis of this harmonic. For example, in contrast to expectation \citep{Wu01}, we find  that the ratio of the amplitudes of the harmonic (2f$_0$) to the  fundamental (f$_0$) depends strongly on wavelength. Figure 6 presents the f$_0$ mode (blue dotted sinusoid) over two periods  along with superpositions (additive combinations) of f$_0$ and 2f$_0$ at different wavelengths. For the pulsations measured at 0.5\,$\upmu$m, the frequencies, amplitudes, and phases are all fit from the 2018 WET time-series photometry. For pulsation frequencies measured at longer wavelengths, the frequencies are fixed at the values determined from the 2018 WET data, but the amplitudes and  phases are fit from the 2018 SpeX synthetic photometry. At 0.5\,$\upmu$m, the superposition of 2f$_0$ and f$_0$ creates a curve that departs slightly from a sinusoid with a slower rise and steeper fall. This contrasts with the prediction of \cite{Wu01} (see also \citealt{Brickhill92a}, figure 7) that such distortion should lead to peaked light curves with sharp ascents and shallow descents.  At 1.51\,$\upmu$m the superposition departs more clearly from a sinusoid and by 2.41\,$\upmu$m the amplitudes of f$_0$ and 2f$_0$ are similar (see also Figure 5). Measuring mode properties for the principal and harmonic frequencies across such a range of wavelengths may offer new constraints on the properties  of surface convection zones or their time-dependent depth changes for these stars.


\begin{figure} 
\vspace{-72pt}
\includegraphics[width=\linewidth]{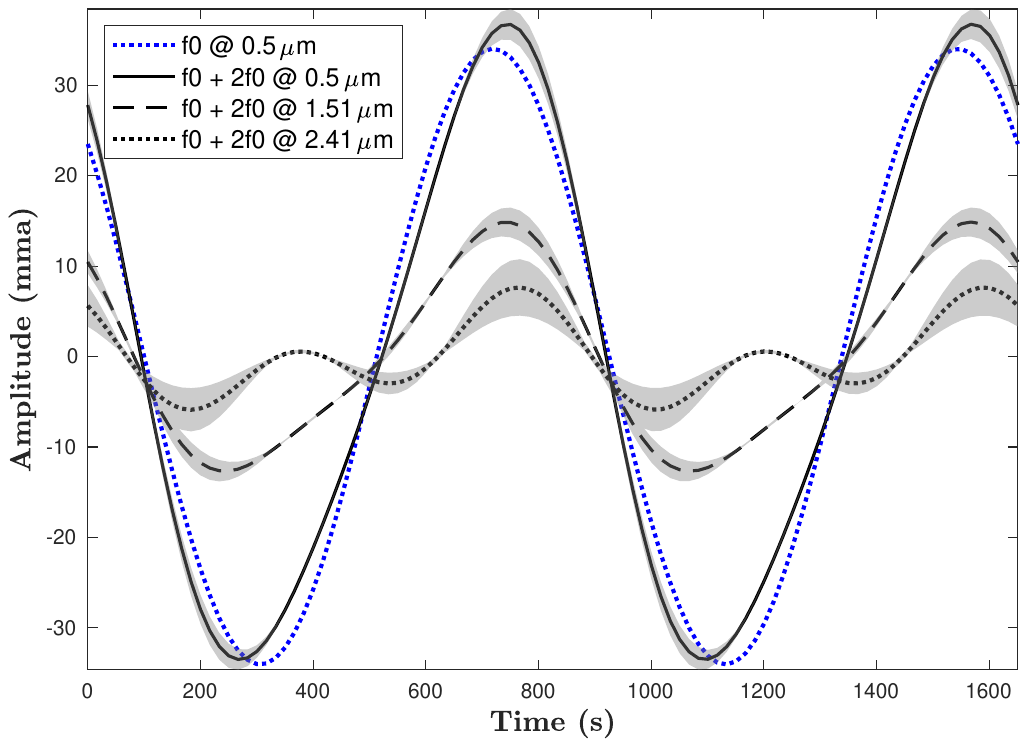}
\vspace{-84pt}
\caption{Comparison of the f$_0$ mode (blue dotted sinusoid) with the superposition of f$_0$ and 2f$_0$ at three different wavelengths (black curves, see legend).  The grey shaded zone covers the $\pm 1 \sigma$ uncertainties in pulsation amplitudes and phases.
\label{fig6}}
\end{figure}

\subsection{Simple Dust Morphology Models}

We note that our observational data extend only to 2.5\,$\upmu$m, where emission from the hottest dust, nearest to sublimation (at the inner edge of a disk or the dust orbital periastron) should peak.  Thus any morphological information that can be derived from the pulsations constrains only the inner edge of the circumstellar environment.  Additionally, at these wavelengths, the dust is only marginally the dominant contributor to the flux \citep[e.g.][]{Xu18}; careful modelling will be required to properly disentangle atmospheric and dust contributions.  With these caveats in mind, we proceed to an initial exploration of how the infrared amplitudes of the combination modes constrain the morphology of the circumstellar dust.  We start by assuming that the dust is more or less in a plane, of some indeterminate thickness.  The dust might be a circular disk, an eccentric disk, or even in a parabolic stream \citep[e.g.][]{Nixon20}.  Given the theories for the origins for the dust, there is no reason to suppose that the rotation axes of the dust and of the white dwarf are in any way aligned.  In considering the illumination of the dust by the white dwarf, the natural coordinates have the $\Theta$ = 0 axis (direction toward the north pole) perpendicular to the dust plane.  In this case, for example, a (2,0) mode on the star (relative to the star’s rotation axis, which is unlikely to be aligned perpendicular to the dust plane) would look to the dust as a collection of the multiplets (2,$-$2), (2,$-$1), (2,0), (2,+1), (2,+2).  Similarly, any $\ell=1$ mode on the star would look to the dust as a collection of the multiplets (1,$-$1), (1,0), and (1,+1).

\subsubsection{Optically Thin Dust}

We consider first the simplest case in which the dust is optically thin at all wavelengths.  In what follows, we continue to work in a coordinate system based on the dust plane.  We consider some simple dust configurations \citep[cf.][]{Cotton20}:
\begin{enumerate}

\item If the dust is distributed uniformly and spherically symmetrically then the only dust response will be to the (0,0) component. This surface brightness distribution of the modulation only occurs as a combination mode. 

\item If the dust is distributed as a uniform circular disk of finite vertical thickness, then by symmetry none of the $m \neq 0$ components produce a net dust response. In addition the (1,0) component produces no net response. Yet the (2,0) component can produce a non-zero net response (see figure~9 in \citealt{Graham90}). In this case, if the underlying basic modes are all $\ell=1$, we expect zero dust response from these, and the only non-zero dust response to be from the combination modes.  But if the underlying modes are all $\ell=2$, then we expect a dust response from both the underlying modes and from the combination modes.

\item If instead the dust is in an eccentric disk (of which the extreme example would be a parabolic stream), then as seen from the star, the dust density will have an $m=1$ (i.e. cos $\Phi$) component. In this case there will also be a response from the underlying $m=+1/-1$ components but not from the underlying $m=+2/-2$ components. Note that in this case there can be a phase difference between stellar pulsations and dust response depending on the geometry.
\end{enumerate}

Thus for these simple dust configurations, one should always expect a response from combination modes (of equal $\ell$), as they always have a (0,0) component. We might expect a response from a (2,0) component. Note that this could be a (2,0) component of a principal pulsation mode or part of the combination response from (1,0)$\times$(1,0) or (2,0)$\times$(2,0) (see appendix of \citealt{Brassard95}). If the dust distribution is circular, then we expect no response to $\ell=1$ components.  But if it is eccentric then there can be a response to such components, with the amplitude dependent on the degree of eccentricity.  

Overall, if we see only a response to combination modes and none to the principal modes, the best guess would be that the underlying modes have $\ell=1$ and the dust distribution is circular.  But, importantly, there are uncertainties here depending on the expected amplitudes of response and the observational uncertainties. For example, the f$_0$, f$_1$, and f$_2$ modes do show amplitudes at 2.5\,$\upmu$m; one would need to be able to disentangle how much of these oscillations at this wavelength might be due to the dust response.  This detailed modeling is beyond the scope of the present paper.

\subsubsection{Optically Thick Dust}

If the dust is optically thick, and indeed that seems probable given the large infrared excess of G29-38, then there are too many possibilities to be considered here.  We restrict the discussion to two obvious simple generalisations, based on previous theoretical ideas and spectral modeling:
\begin{enumerate}

\item Consider the dust to be arranged in a circular, infinitesimally thin disk, which is optically thick in the vertical direction. In this case none of the $m \neq 0$ components contribute to the infrared response. But both the (1,0) and the (2,0) components would give rise to a response.

\item Consider the dust to be in a circular disk of finite thickness so that the main dust response comes from the inner edge of the disk \citep[cf.][]{Ballering22}. In this case, assuming an intermediate inclination angle, one expects no contribution from (1,0), but contributions from (2,0) as well as from the $m \neq 0$ components, but not from $m = +2/-2$ components.
\end{enumerate}

While our observations of higher infrared contributions in the combination modes may favor some of these models, in order to definitively rule out any of them requires modeling the variations of amplitudes and phases as functions of wavelength.

\section{Conclusions}
\label{conclusions}

We obtained time-series near-infrared spectrophotometry of G29-38 using SpeX at the IRTF in 2018 and 2020, along with contemporaneous optical WET time-series photometry.  We detected six principal pulsation modes, one harmonic, and three combination modes in the 2018 dataset.  Pulsations during the 2020 observations were too weak to see clear trends in amplitude as a function of wavelength.  Among the 2018 pulsations with sufficient S/N to measure across the full range of observed wavelengths, the three principal modes all showed declining pulsation amplitudes with increasing wavelength.  These pulsation amplitudes decreased from 20 to 35\,mma in $V$, $g$, and other WET optical bands to $\approx$5\,mma at 2.5\,$\upmu$m.  The three nonlinear combination modes from 2018 exhibited strikingly different behavior, rising from approximately 6 to 9\,mma in the optical to 8 to 19\,mma in the near-infrared.  The harmonic of the principal f$_0$ mode may show similar behavior, though at a reduced level.

The basic theory of the origin and the behavior of harmonics and combination modes for stars with a surface convection zone, which includes all ZZ Ceti stars, has been set out by \citet{Brickhill92a, Brickhill92b} and by \citet{Wu01}. We have found some discrepancies with the expectations laid out in those papers. These papers predict for a single mode that the ratio of the amplitude of the first harmonic to the amplitude of the underlying mode should be independent of wavelength and that the effect of the first harmonic is to give rise to a peaked light curve with a sharp ascent and shallow descent. It is evident from Figures 5 and 6 that our results disagree.  Also, these papers predict that the ratios of the amplitudes of the principal modes to the amplitudes of their resultant combination modes are independent of wavelength. While we have not performed the modeling to separate the dust vs.\ photospheric contributions, from Figure 6 and the time-averaged spectral energy distribution modeling of \citet{Reach09}, it appears that our results disagree. These findings warrant further investigation.

We conclude that the dominant near-infrared response is most likely due to the isotropic component of the combination modes.  Unfortunately, isotropic pulsation components are of no use in determining the geometric structure of the dust.  To make future gains in determining the dust morphology, we need to measure the dust response to modes with $\ell \geq 1$, i.e.\ principal modes and perhaps non-isotropic components of the combination modes.  This will require careful modeling.  Additionally, in order to clarify what fraction of the infrared pulsation is from the dust response versus from the stellar photosphere, we need detailed modeling of atmospheric pulsation modes, which depend on identifying the spherical harmonics of these modes.  An alternative approach would be to observe at wavelengths longer than $5\,\upmu$m, where G29-38's spectrum is almost entirely due to the heated dust \citep{Reach09}.  Because of the time-varying nature of the stellar pulsation modes, some epochs have only weak pulsations that are insufficient to constrain the dust environment.  However, whenever G29-38 exhibits different pulsation frequencies these may provide new diagnostics and the dust distribution itself may vary with time. The phase behavior of the pulsation frequencies also  warrants further investigation and may help resolve dust and surface pulsation geometries.

While not a goal of this study, our analysis of the combination modes likely indicates that the observed principal modes all have the same $\ell$ value.  This constraint is both a prediction and a potentially useful tool for future mode identification on this important star.

\section*{Acknowledgements}
Data in this paper are based on observations obtained at the Infrared Telescope Facility, which is operated by the University of Hawaii under contract 80HQTR19D0030 with the National Aeronautics and Space Administration.  Supporting observations were obtained at the facilities listed in Tables 2 and 3.  This material is based upon work supported by the National Science Foundation under Grant No.\ AST-1715718.  This research is funded by the Science Committee of the Ministry of Science and Higher Education of the Republic of Kazakhstan (Grant No. BR20280974).  EP acknowledges the Europlanet 2024 Research Infrastructure project funded by the European Union's Horizon 2020 Research and Innovation Programme (Grant agreement No. 871149).

\bibliographystyle{mnras}
\bibliography{WD_generic} 


\label{lastpage}
\end{document}